\documentclass[10pt,english,aps,amssymb,prl,twocolumn, showpacs]{revtex4-1}
\pdfoutput=1
\usepackage[T1]{fontenc}
\usepackage[latin9]{inputenc}
\usepackage{amsmath}
\usepackage{graphicx}
\usepackage{amssymb}
\usepackage{mathtools}
\usepackage{esint}
\usepackage{verbatim}
\usepackage{bm}
\usepackage{hyperref}
\usepackage{amsfonts}
\usepackage{bbm}
\usepackage{wrapfig}
\usepackage{color}
\usepackage[top=2cm, bottom=2cm, left=2cm, right=2cm]{geometry}
\makeatletter
\@ifundefined{textcolor}{}
{
 \definecolor{WHITE}{gray}{1}
 \definecolor{RED}{rgb}{1,0,0}
 \definecolor{GREEN}{rgb}{0,1,0}
 \definecolor{BLUE}{rgb}{0,0,1}
 \definecolor{CYAN}{cmyk}{1,0,0,0}
 \definecolor{MAGENTA}{cmyk}{0,1,0,0}
 \definecolor{YELLOW}{cmyk}{0,0,1,0}
 }

\renewcommand{\phi}{\varphi}
\renewcommand{\epsilon}{\varepsilon}

\renewcommand{\vec}[1]{{\bf #1}}
\renewcommand{\vr}[1]{{\bf #1}}
\renewcommand{\kappa}{\varkappa}

\renewcommand{\vr}[1]{{\mathbf #1}}

\newcommand{\intt}{\int\!}
\renewcommand{\phi}{\varphi}
\renewcommand{\epsilon}{\varepsilon}

\renewcommand{\vec}[1]{{\mathbf #1}}

\usepackage{babel}

\begin{document}
\title {Topological state engineering by potential impurities on chiral superconductors }
\author{Vardan Kaladzhyan$^{1,2}$ }
\author{Joel R\"ontynen$^3$}
\author{ Pascal Simon$^2$}
\author{Teemu Ojanen$^3$}
\affiliation{$^1$Institut de Physique Th\'eorique, CEA/Saclay, Orme des Merisiers, 91190 Gif-sur-Yvette Cedex, France }
\affiliation{$^2$Laboratoire de Physique des Solides, CNRS, Univ. Paris-Sud,
Universit\'e Paris-Saclay, 91405 Orsay Cedex, France }

\email[Correspondence to ]{teemuo@boojum.hut.fi}
\affiliation{$^3$Department of Applied Physics (LTL), Aalto University, P.~O.~Box 15100,
FI-00076 AALTO, Finland }
\date{\today}
\begin{abstract}
In this work we consider the influence of potential impurities deposited on top of two-dimensional chiral superconductors.  As discovered recently,  magnetic impurity lattices on an $s$-wave superconductor may give rise to a rich topological phase diagram. We show that similar mechanism takes place in chiral superconductors decorated by non-magnetic impurities, thus avoiding the delicate issue of magnetic ordering of adatoms. We illustrate the method by presenting the theory of potential impurity lattices embedded on chiral $p$-wave superconductors. While a prerequisite for the topological state engineering is a chiral superconductor, the proposed procedure results in vistas of nontrivial descendant phases with different Chern numbers.  
       
\end{abstract}
\pacs{73.63.Nm,74.50.+r,74.78.Na,74.78.Fk}
\maketitle
\bigskip{}

\emph{Introduction--} Engineering novel quantum phases of matter with exotic properties is a rapidly growing trend in contemporary physics.  The main goal is to employ simpler and well-understood ingredients and methods to create more complex structures with desirable properties.  Recent promising efforts to realize \cite{mourik,das,albrecht} topological superconductivity in nanowire systems \cite{lutchyn, oreg} demonstrate the power of the approach. While it seems unlikely that Nature directly provides us  with Majorana quasiparticles that could be employed in quantum information applications \cite{nayak}, it is increasingly probable that those can be achieved in laboratory.  In the spirit of engineering novel controllable states of matter, we show how to realize a complex hierarchy of topological phases with potential impurity superstructures adsorbed on chiral superconductors.

Magnetic atoms on $s$-wave superconductors give rise to Yu-Shiba-Rusinov subgap states \cite{yu,shiba,rusinov,balatsky} which have been probed experimentally by scanning tunneling microscopy (STM) \cite{yaz,shuai,franke,menard}. Superstructures fabricated from magnetic atoms  are currently under active experimental \cite{np2,pawlak,ruby} and theoretical research \cite{choy,np,brau1,klin,vazifeh,pientka2,pientka3,poyh,ront,heimes,reis,bry,heimes2,west,brau2,poyh2}. Intriguing properties of these systems include possibility for various one dimensional (1D) topological superconducting phases with Majorana bound states and rich 2D topological phases \cite{ront1,ront2,li1,nakosai}. Topologically nontrivial phase is known to arise in 1D ferromagnetic arrays when the underlying superconductor has a strong Rasha spin-orbit coupling or in arrays with helical magnetic textures. In 1D structures there are theoretical arguments why magnetic self-tuning could result in a nontrivial ground state \cite{brau1,klin,vazifeh,reis,brau2,schecter}, though in real systems there are number of complications.  In particular, in 2D structures the nature and tuneability of magnetic textures is a delicate and largely unsolved question. 
\begin{figure}
\includegraphics[width=0.99\columnwidth]{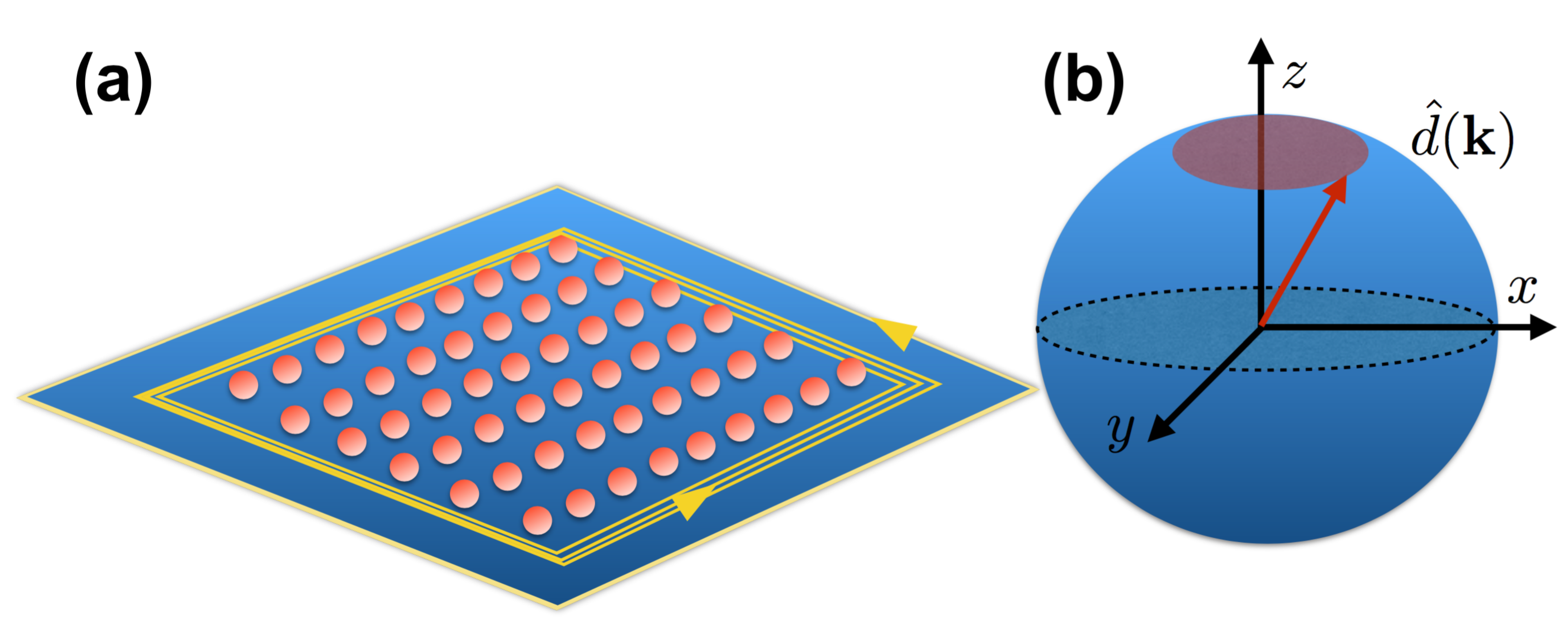} 
\caption{(a) Schematic representation of the studied system, consisting of potential scatterers deposited on top of a chiral superconductor. The topological phase on the impurity lattice can be be widely modified from that of the underlying compound. (b) In two-band models the Chern number can be illustrated through the motion of the $\hat{d}(\vec{k})$ vector on the unit sphere. A long-range hopping translates to high Chern numbers through rapid rotation of $\hat{d}(\vec{k})$. }
\label{first}
\end{figure}

Very recently it was proposed that potential impurities could be utilized to realize interesting topological states in 1D structures \cite{neupert} and 2D toy models \cite{kimme}. The procedure requires a non-$s$-wave superconductor host material with chiral or helical pairing components but circumvents the need for specific magnetic textures of adatoms. In the present work we provide a microscopic theory of potential impurity structures on chiral superconductors. We show that given a nontrivial chiral superconductor, the potential impurities give rise to a complex hierarchy of distinct nontrivial phases. The Chern number of the phase can be structurally designed by employing different impurities and varying the impurity lattice constant. We illustrate the procedure with a chiral $p$-wave superconductor. However, our results are not restricted to chiral $p$-wave systems and also apply to time-reversal breaking $s+p$-mixtures, higher chiral superconductors and the artificial $p$-wave model realized in sandwich structures of a 2D semiconductor proximity coupled to an $s$-wave superconductor and a ferromagnetic insulator.

\emph{Chiral $p$-wave systems--} Here we formulate the theory describing the system in Fig.~\ref{first} a). The bulk electrons in a 2D  spinless $p_x+ip_y$ superconductor are described by a Bogoliubov-de Gennes (BdG) Hamiltonian
\begin{equation*}
\mathcal{H}_\vec{p}^{(\rm bulk)} =  \xi_\vec{p}\tau_z + \kappa\left( p_x\tau_x - p_y\tau_y \right),
\end{equation*}
expressed in the Nambu basis $(\hat{\Psi}_{p}, \hat{\Psi}^\dagger_{-p})$. Here the single-particle energy is $\xi_\vr{p} = \frac{p^2}{2m} - \epsilon_F$ with the Fermi energy $\epsilon_F$, and $\kappa$ is the superconducting $p$-wave pairing amplitude which is taken as real and positive. The Pauli matrices $\tau_i$ operate in the particle-hole space. The collection of adatoms act as local potentials described by
\begin{equation*}
\mathcal{H}^{(\rm imp)}(\vec{r}) = U\tau_z\sum_n\,\delta(\vec{r}-\vec{r}_n),
\end{equation*}
where $\vr{r}_n$ are the positions of the atoms and $U$ is the impurity strength. Our treatment is also valid if we consider impurities of a finite size (see the supplementary information (SM) \cite{sm} for details ). The total Hamiltonian consists of the sum $\mathcal{H} = \mathcal{H}^{(\rm bulk)} + \mathcal{H}^{(\rm imp)}$.

Each potential impurity atom binds a single physical subgap state \cite{kaladzhyan2015}, which in the BdG formalism is represented by a pair of states at energies $\epsilon=\pm\frac{\gamma\beta^2-\sqrt{1+\beta^2(1-\gamma^2)}}{1+\beta^2}\Delta_t$ for repulsive potential $\beta>0$. For attractive potential $\beta<0$ the solutions are otherwise the same with the exception of a minus sign in front of the square root \cite{sm}. Here we have defined quantities $\beta=\pi\nu U$, 
$\gamma=\frac{\tilde{\kappa}}{\sqrt{1+\tilde{\kappa}^2}}$, $\tilde{\kappa}=\frac{\kappa}{v_F}$ and $\Delta_t= \frac{\kappa k_F}{\sqrt{1+\tilde{\kappa}^2}}$, where $v_F$ is the Fermi velocity and $\nu$ the density of states in the bulk \cite{footnote}.

The parameter $\Delta_t$ represents the $p$-wave bulk gap determining the coherence length $\xi^{-1}=\frac{\Delta_t}{v_F}$ and $\beta$ is a dimensionless impurity strength. Strong impurities with $\beta\gg 1$ give rise to deep-lying subgap states close to the Fermi level while weak impurity states reside near the gap edge. Analogous to the Shiba states in a 2D systems \cite{menard}, the potential impurity wave functions have asymptotic form $e^{ik_Fr-r/\xi_E}/\sqrt{k_Fr}$ away from the impurity where the decay length is given by $\xi_{E}=\xi/\sqrt{1-(E/\Delta_t)^2}$.

When impurity atoms are arranged into a regular array with a lattice constant $a<\xi$, the impurity states bound to a particular atom are hybridized with several nearest neighbours.  This leads to the formation of subgap energy bands which support rich topological properties. To study the topological properties of the subgap bands, we formulate effective low-energy theory valid in the deep-dilute impurity regime $\beta\gg1$, $\sqrt{k_Fa}\gg1$ in the vicinity of the Fermi level. However, as we discussed below, the effective theory yields an exact topological phase diagram which is valid also outside the deep-dilute regime.  As outlined in the supplementary \cite{sm}, a similar procedure that was applied in the Shiba systems \cite{pientka2, pientka3, bry, ront1} results in a description of the impurity lattice in terms of the tight-binding Hamiltonian
\begin{equation}\label{h1}
\begin{gathered}
H_{mn} = \begin{pmatrix} h_{mn} & \Delta_{mn} \\ (\Delta_{mn})^\dagger & -h_{mn}^* \end{pmatrix}.
\end{gathered}
\end{equation}
The effective Hamiltonian has an $N\times N$ BdG block structure, where $N$ is the number of impurity atoms.  The BdG blocks are given by
\begin{equation}\label{h2}
\begin{split}
h_{mn} &= \begin{dcases} \epsilon_0& m=n \\
 A(r_{mn})& m\neq n \end{dcases}
\\
\Delta_{mn} &= \begin{dcases} 0 & m=n \\
B(r_{mn}) \frac{x_{mn} + i y_{mn}}{r_{mn}} & m\neq n, \end{dcases}
\end{split}
\end{equation}
where the onsite term $ \epsilon_0=\Delta_t(\gamma-\beta^{-1})$ arises from the decoupled impurity energy,  $r_{mn} = |\vr{r}_m-\vr{r}_n|$ is the distance between two impurity lattice sites and $x_{mn} = x_m - x_n$, $y_{mn} = y_m - y_n$.  The matrix elements depend on the functions  
\begin{equation*}
\begin{split}
A(r) &= - \frac{2\Delta_t}{\pi}\text{Re} \Big\{ \eta K_0[-i\eta k_Fr] \Big\},
\\
B(r) &=-i\frac{2\Delta_t}{\pi} \text{Re} \Big\{ \eta K_1[-i\eta k_Fr] \Big\},
\end{split}
\end{equation*}
where $K_i(x)$ stands for the modified Bessel function of the second kind with index $i$ and $\eta=1+i\tilde{\kappa}$.  The block matrices in Eq.~($\ref{h2}$) define a hopping model where the amplitudes satisfy asymptotic behaviour  $\Delta_{mn}, h_{mn}\sim\frac{e^{-r_{mn}/\xi}}{\sqrt{r_{mn}}}$ at long distances. The model (\ref{h1}) with entries (\ref{h2}) is a lattice discretized chiral superconductor with rich topological properties discussed below.

\begin{figure}
\includegraphics[width=0.99\columnwidth]{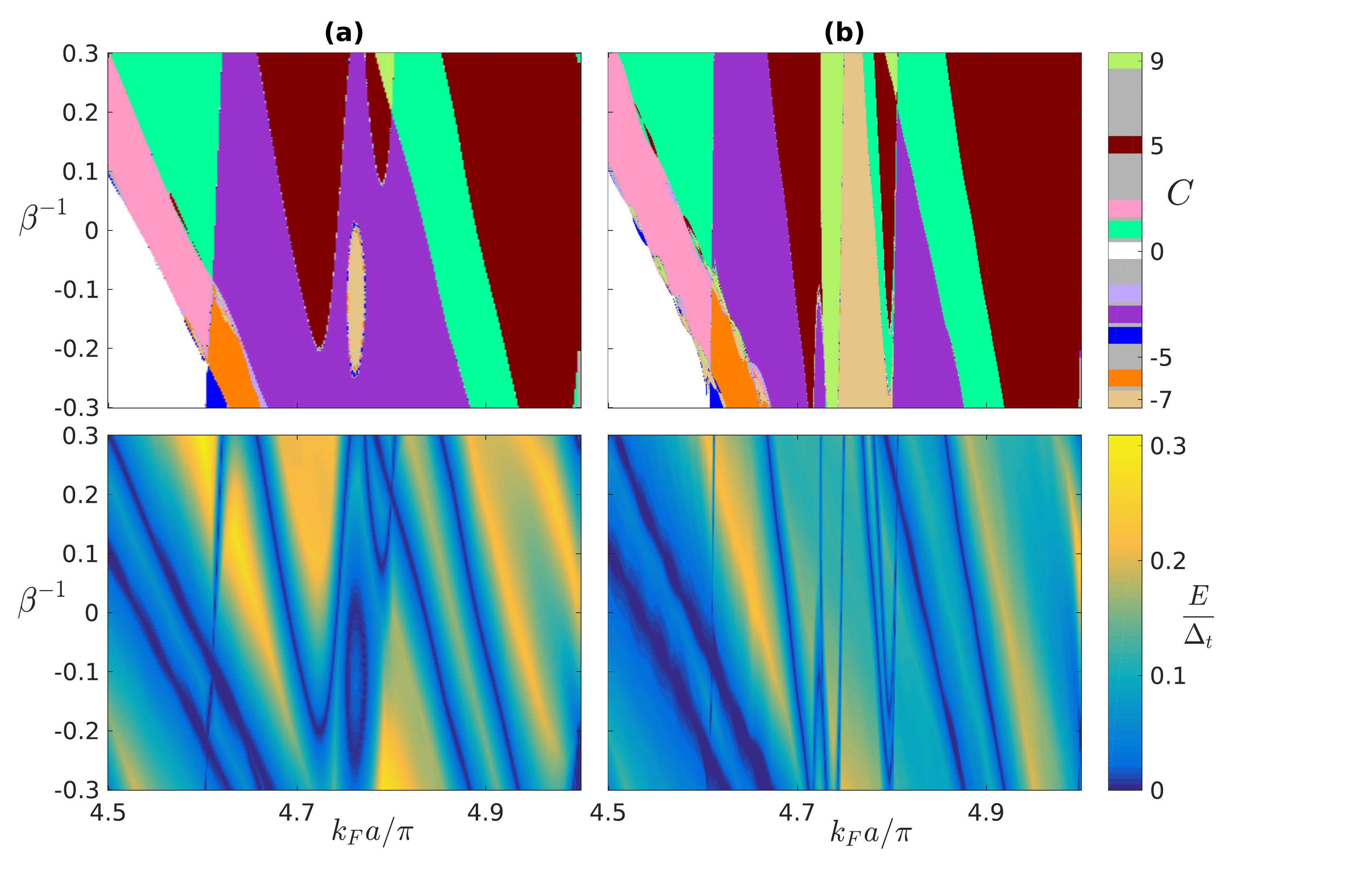} 
\caption{(a): Chern number (above) and energy gap (below) diagrams for a square lattice of impurities with lattice constant $a$ and coherence length $\xi/a=5$. The quantity $\beta^{-1}$ in the vertical axis controls the strengths of the impurity. The $\beta^{-1}=0$ line, corresponding to infinite impurity potential $|U|=\infty$, divides the repulsive and attractive impurity regions. The horizontal axis $k_Fa$ controls the hybridization between the bound states centred at different impurity sites. (b): Same as (a) but for coherence length $\xi/a=10$.}
\label{one}
\end{figure}

\emph{Topological properties--} The topological phase diagram of the effective model (\ref{h2}) is  conveniently extracted in momentum space. For any Bravais lattice we can define  Fourier transforms
\begin{equation*}
\begin{split}
d_x(\vr{k}) &= \text{Re} \sum_{\vr{R}} e^{i\vr{k}\cdot\vr{R}} \Delta_{\vr{R}},\quad
d_y(\vr{k}) = - \text{Im} \sum_{\vr{R}} e^{i\vr{k}\cdot\vr{R}} \Delta_{\vr{R}},
\\
d_z(\vr{k}) &= \sum_{\vr{R}} e^{i\vr{k}\cdot\vr{R}} h_{\vr{R}},
\end{split}
\end{equation*}
where the sum is over all the lattice vectors $\vr{R}=(x_{mn}, y_{mn})$. The Hamiltonian can then be written in a simple form $H(\vr{k}) = \vr{d}(\vr{k})\cdot\boldsymbol{\sigma}$ with energies $E(\vr{k}) = \pm |\vr{d}(\vr{k})|$. The effective Hamiltonian $H(\vr{k})$ describes gapped two-band model satisfying the particle-hole symmetry $\mathcal{C}H(\vec{k})^*\mathcal{C}^{-1}=-H(-\vec{k})$, where $\mathcal{C}=\sigma_x\mathcal{K}$ and $\mathcal{K}$ denotes complex conjugation. The studied model belongs to the Altland-Zirnbauer class $D$, admitting a $\mathbb Z$-valued classification by Chern numbers \cite{schnyder}.  For two-band models the Chern number is found by  evaluating the expression 
\begin{equation}\label{C1}
C = \frac{1}{4\pi} \smashoperator{\int_{\rm BZ}}\! d^2k\, \frac{\vr{d}}{|\vr{d}|^3} \cdot \bigg( \frac{\partial\vr{d}}{\partial k_1} \times \frac{\partial\vr{d}}{\partial k_2} \bigg),
\end{equation}
which yields integers. The integer value of the Chern number can be visualized through construction depicted in Fig.~\ref{first} (b). The Hamiltonian defines a unit vector $\hat{d}(\vec{k})=\vec{d}(\vec{k})/|\vec{d}(\vec{k})|$ which can be depicted as a point on the surface of a unit sphere. Absolute value of the Chern number measures how many times $\hat{d}(\vec{k})$ covers the sphere when $\vec{k}=(k_x,k_y)$ covers the Brillouin zone of the impurity lattice. The long-range hopping gives rise to rapidly rotating components of $\hat{d}$ vector and thus may lead to chiral states with Chern numbers much larger than unity.

As pointed out in the SM \cite{sm}, the effective description (\ref{h1}), derived under assumptions of  a deep and dilute impurity configuration $\beta\gg1$, $\sqrt{k_Fa}\gg1$, actually acts as a topological Hamiltonian yielding the \emph{exact} phase diagram which is also valid outside the deep-dilute regime. This happens because at the topological phase transition, accompanied by the energy gap closing, the effective model  (\ref{h1}) becomes exact irrespectively of the values of $\beta$ and $k_Fa$.

\begin{figure}
\includegraphics[width=0.99\columnwidth]{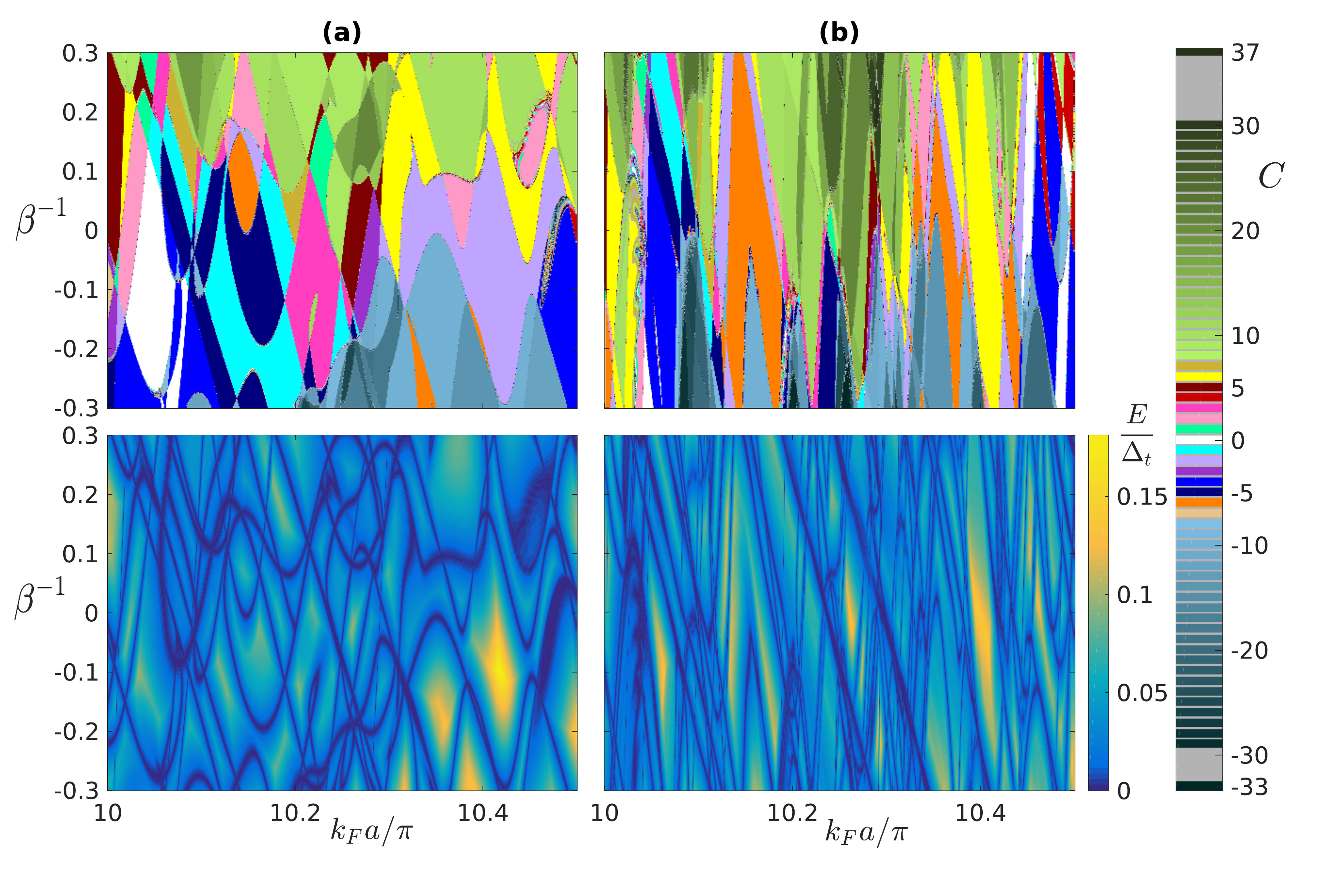} 
\caption{The same quantities as in Fig.~\ref{one} but for larger values of the hybridization parameter $k_Fa$. }\label{two}
\end{figure}
\begin{figure}
\includegraphics[width=0.99\columnwidth]{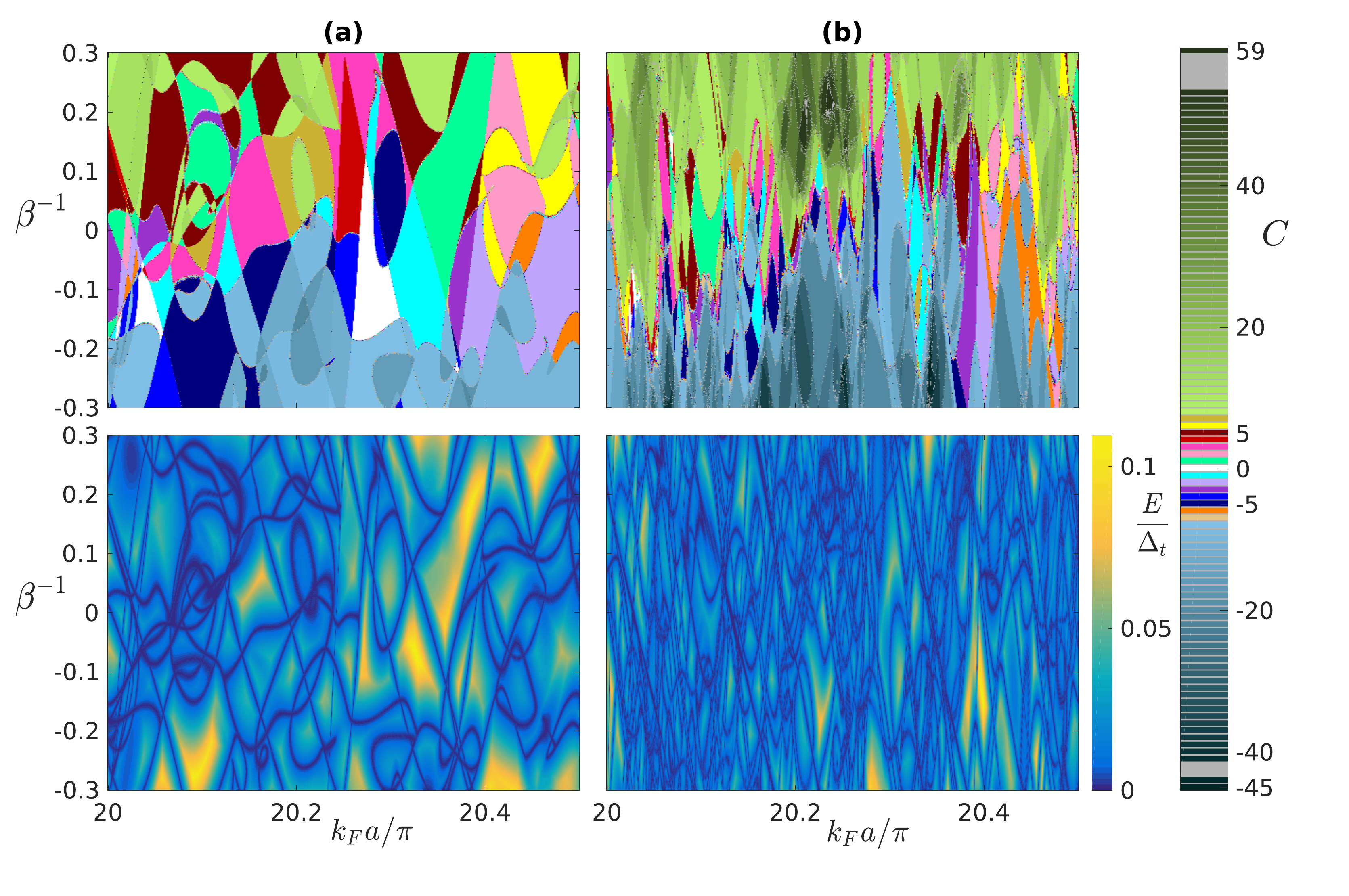} 
\caption{The same quantities as in Fig.~\ref{one} but for larger values of the hybridization parameter $k_Fa$. }
\label{three}
\end{figure}

In Fig.~\ref{one} we have plotted the topological phase diagram and the energy gap diagram for square lattices. It is clearly evident that the system possesses multiple phases which can be tuned by the separation and strength of the impurities. For higher values of the hybridization parameter $k_Fa$ the hopping is highly oscillatory, thus leading to more rapid alternation of  Figs.~\ref{two} and \ref{three}. The generic features of the phase diagrams seem to be be in line with the Chern mosaic behaviour discovered in magnetic lattices \cite{ront1,ront2}. For robust states the energy gaps are of the order of $0.1-0.2\Delta_t$. Probably larger gaps can be obtained, but studying those would require more elaborate theory as the employed approximations become unreliable. Potential impurity superstructures clearly allow remarkable possibilities for topological state engineering in the studied system without uncertainty associated to the magnetic textures. 
  
We have also diagonalized the system on an infinite strip geometry, where the topological edge modes show up as states traversing the bulk gap. These results are discussed in  more detail in the supplement.

\emph{Physical realizations--} In the above we have considered potential impurities in spinless chiral $p$-wave superconductors. Our theory can be straightforwardly generalized to the candidate state of Sr$_2$RuO$_4$ where  the opposite spins pair to form $L_z=1$ Cooper pairs. Since potential impurities do not mix spin,  the $4\times4$ model with spin leads to two identical but decoupled $2\times 2$ blocks of form Eq.~(\ref{h1}). The Chern number can be evaluated for each block separately, leading to doubling of the Chern number and the edge modes compared to the spinless case.  

However, there are various other candidates for the host materials. The requirements for topological state engineering by potential impurities are rather general and met in a variety of other systems as well. The basic ingredient is that localized potentials must bind subgap bound states in the host material. These bound states in chiral superconductors are generic since Anderson's theorem which guarantees the robustness of $s$-wave superconductors to potential disorder \cite{balatsky} is not operational in time-reversal breaking systems. The second requirement is the phase winding structure $\Delta_0e^{in\phi_k}$, where $\mathrm{tan} \,\phi_k=k_y/k_x$, of the gap function of the unperturbed bulk. This will translate to a type of $(x_{ij}\pm iy_{ij})^n/r_{ij}$ phase structure of the gap function $\Delta_{ij}$ in the effective low-energy BdG Hamiltonian (\ref{h2}), indicating topologically nontrivial superconductivity. In addition, algebraically decaying hopping up to the coherence length is also a universal feature of gapped states. Thus any 2D chiral ($p$-, $d$-, $f$...-wave) superconductor satisfies the general requirements and exhibit the characteristic features of the studied chiral $p$-wave model. We note that different crystal structures of the bulk give rise to distinct lattice regulations of chiral gap functions. Also, for a continuum expression $\Delta(\vec{k})\sim(k_x+ik_y)^n$, corresponding to Chern number $n$, there exists many different lattice versions. However, in the case where the impurity lattice constant is much larger than that of the underlying superconductor, the continuum approximation should prove sufficient. 

Dominantly $p$-wave superconductors with $s$-wave pairing amplitude, having a gap structure $\Delta_s+\Delta_pe^{i\phi_k}$, is also a sufficient starting point for topological state engineering when $\Delta_p>\Delta_s$. In this case potential-impurity induced bound states exist \cite{kaladzhyan2015} and phase winding is inherited to the effective low-energy model. Such $s+p$-wave structure is satisfied in the artificial chiral superconductor realized in 2D Rashba-coupled semiconductors sandwiched by an $s$-wave superconductor and a ferromagnetic insulator \cite{sau} at sufficiently strong magnetization. Patterning the semiconductor layer with potential impurities or otherwise realizing the potential lattice by applying an external structured potential gate would enable fabrication of nontrivial topological states far beyond Chern number $|C|=1$. 

Chiral and time-reversal breaking superconductors have also been predicted in various other low-dimensional systems. While these have not been observed in experiments so far, it is plausible that some will be realized in the future. At that point a large number of other chiral states will immediately become accessible through topological state engineering by potential superstructures.

 \emph{Discussion--}
The bulk topology in a topologically nontrivial state is reflected on its boundary properties. This property could be employed in experimental identification of nontrivial bulk states. The subgap density of states in chiral superconductors arises due to the chiral edge modes as illustrtaed in Fig.~1 (a). Probing the local density of states by Scanning Tunneling Microscopy (STM) reveals that the subgap modes are localized on the boundary of the impurity lattice \cite{ront1}. This method can be employed to show that the impurity lattice is in a different topological phase than the underlying chiral superconductor. Experimental extraction of specific value of the Chern number of a superconductor, while in principle possible, is an unsolved issue at present. However, by fabricating interfaces between lattices of, say, different lattice constants it is possible to compare whether the two adjacent structures belong to the same topological phase. If the structures belong to different phases, there must exist pronounced subgap local density of states at the boundary due to topological edge modes.

The circulating Majorana edge modes, depicted in Fig.~1 (a), carry heat in otherwise gapped systems and could find applications in the future electronics as chiral heat guides. These waveguides could be designed on top of the superconductor by employing different impurity lattice structures. The Chern number of lattice yields the number of parallel thermal edge channels, so high Chern number states are generally more effective thermal conductors compared to low Chern number states. Also, Majorana bound states trapped in lattice defects could also be interesting from quantum information point of view. While the applications of chiral superconductors are still emerging,  our work points to a conceptually simple method to obtain them in nanofabricated structures.

\emph{Conclusions--}
In this work we proposed a method to engineer  topological states by potential impurities deposited on 2D chiral superconductors. In particular, we presented a microscopic theory of chiral $p$-wave superconductors with impurity lattices. This allowed us to calculate the topological phase diagram for general impurity strengths and hybridization. Our results have remarkable conceptual and practical consequences: given a 2D chiral superconductor, it is possible to fabricate a large number of nontrivial descendant states by a straightforward procedure. Because potential-induced subgap states are generic in time-reversal breaking superconductors and superfluids, our results have universal appeal irrespective of the platform and microscopic details of the chiral state.

The authors acknowledge Aalto Science-IT project for the computational resources and the Academy of Finland (T. O.) and the Finnish Cultural Foundation (J. R.) for support. P.S. would like to acknowledge the financial support from the French Agence Nationale de la Recherche through the contract  Mistral.

\appendix
\numberwithin{equation}{subsection}

\widetext
\section*{Supplemental  Information- derivation of the effective Hamiltonian for chiral $p$-wave model}

\subsection*{Spinless $p$-wave case}

The Bogoliubov-de Gennes Hamiltonian can be separated into a bulk and an impurity term, $\mathcal{H} = \mathcal{H}^{(\rm bulk)} + \mathcal{H}^{(\rm imp)}$, where the bulk Hamitonian,
\begin{equation*}
\begin{gathered}
\mathcal{H}_\vec{p}^{(\rm bulk)} =  \xi_\vec{p}\tau_z + \kappa\left( p_x\tau_x - p_y\tau_y \right),
\end{gathered}
\end{equation*}
in the Nambu basis $(\hat{\Psi}_{p}, \hat{\Psi}^\dagger_{-p})^T$. Here $\xi_\vr{p} = \frac{p^2}{2m} - \epsilon_F$ with the Fermi energy $\epsilon_F$, and $\kappa$ is the superconducting $p$-wave pairing amplitude. The Pauli matrices $\tau_i$  operate in the particle-hole. The impurity Hamiltonian,
\begin{equation*}
\begin{gathered}
\mathcal{H}^{(\rm imp)}(\vr{r}) =  U\tau_z\sum_n\,\delta(\vec{r}-\vec{r}_n)
\end{gathered}
\end{equation*}
describes an potential impurity of strength $U$ located at position $\vr{r}_j$. 

The BdG equation $\mathcal{H}(\vr{r}) \Psi(\vr{r}) = E \Psi(\vr{r})$ yields
\begin{equation*}
\begin{gathered}
\big[ E - \mathcal{H}^{(\rm bulk)}(\vr{r}) \big] \Psi(\vr{r}) 
= U\tau_z\sum_n\,\delta(\vec{r}-\vec{r}_n)\Psi(\vr{r}).
\end{gathered}
\end{equation*}
We change to momentum space using the Fourier transform $\Psi(\vr{r}) = \intt \frac{d\vr{p}}{(2\pi)^2} e^{i\vr{p}\cdot\vr{r}} \Psi_\vr{p}$ and thus obtain
\begin{equation*}
\begin{gathered}
\big[ E - \mathcal{H}^{(\rm bulk)}_\vr{p} \big] \Psi_\vr{p} = U\tau_z\sum_j \, e^{-i\vr{p}\cdot\vr{r}_j} \Psi(\vr{r}_j).
\end{gathered}
\end{equation*}
Solving for $\Psi_\vr{p}$ and going back to real space, we thus find
\begin{equation}
\begin{gathered}
\Psi(\vr{r}) = U\sum_j G_0(E,\vr{r}-\vr{r}_j)\, \tau_z\, \Psi(\vr{r}_j),
\label{impurityeq}\tag{A.1}
\end{gathered}
\end{equation}
where 
$G_0(E,\vr{r}) = \intt \frac{d\vr{p}}{(2\pi)^2} e^{i\vr{p}\cdot\vr{r}} \big[ E - \mathcal{H}^{(\rm bulk)}_\vr{p} \big]^{-1} $
is the bulk Green function.

\subsection*{Single potential impurity bound states}
Let us first consider a single impurity at the origin. We can set $\vr{r}=\vr{0}$ in Eq.\ (\ref{impurityeq}) and obtain the following eigenvalue equation for the impurity energies:
\begin{equation*}
\begin{gathered}
\big[\mathbbm{1} - U G_0(E,\vr{0})\, \tau_z\big] \Psi(\vr{0}) = 0.
\end{gathered}
\end{equation*}
As we are considering only subgap energies, we can evaluate $G_0(E,\vr{0})$ assuming $|E|<\Delta_t$, yielding $G_0(E,\vr{0})=-\frac{\pi\nu}{\sqrt{1+\tilde{\kappa}^2}}\frac{\Delta_t}{\sqrt{\Delta_t^2-E^2}}\big[E\mathbbm{1}-\gamma \tau_z \big]$. Here we have defined quantities $\gamma=\frac{\tilde{\kappa}}{\sqrt{1+\tilde{\kappa}^2}}$, $\tilde{\kappa}=\frac{\kappa}{v_F}$ and $\Delta_t=\frac{\kappa k_F}{\sqrt{1+\tilde{\kappa}^2}}$ where $v_F$ is the Fermi velocity and $\nu$ the density of states in the bulk.  Inserting this result into the single-impurity eigenvalue equation yields
\begin{equation}
\begin{gathered}
\Big[\mathbbm{1} + \frac{\beta}{\sqrt{\Delta_t^2-E^2}} (E\tau_z -\gamma \Delta_t \mathbbm{1})\Big] \Psi(\vr{0}) = 0,
\label{singleimp}\tag{A.2}
\end{gathered}
\end{equation}
where $\beta=\pi\nu U$ is the dimensionless impurity strength determining the bound state energy. Eq.~ (\ref{singleimp}) has two solutions for both repulsive $\beta>0$ and attractive $\beta<0$ impurities. For $\beta>0$ the solutions have energies $ E= \epsilon_{\pm}=\pm\frac{\gamma\beta^2-\sqrt{1+\beta^2(1-\gamma^2)}}{1+\beta^2}\Delta_t$ with eigenstates $\Psi_+(\vr{0})=(1,0)^T$ and $\Psi_-(\vr{0})=(0,1)^T$. For  $\beta<0$ the energies are  $E= \epsilon'_{\pm}=\pm\frac{\gamma\beta^2+\sqrt{1-\beta^2(1-\gamma^2)}}{1+\beta^2}\Delta_t$. 

\subsection*{Lattice of potential impurities}
In case of multiple impurities at positions $\vr{r}_i$, Eq.\ (\ref{impurityeq}) becomes
\begin{equation}\label{exact}\tag{A.3}
\begin{gathered}
\big[\mathbbm{1} - UG_0(E,\vr{0})\,\tau_z\big] \Psi(\vr{r}_i) = U\sum_{j\neq i} G_0(E,\vr{r}_i-\vr{r}_j)\,\tau_z \Psi(\vr{r}_j).
\end{gathered}
\end{equation}
To proceed, we need to evaluate $G_0(E,\vr{r})$ for $|\vr{r}|>0$. Employing methods outlined in Ref.~\cite{kaladzhyan2016} we obtain
\begin{equation*}
\begin{gathered}
 G_0(E,\vr{r})=\begin{pmatrix} EX_0(\vr{r})+X_1(\vr{r})& iX_2^+(\vr{r})\\ -iX_2^-(\vr{r}) &  EX_0(\vr{r})-X_1(\vr{r}) \end{pmatrix},
\end{gathered}
\end{equation*}
where 
\begin{equation*}
\begin{split}
X_0(\vr{r}) &=  - \intt \frac{d\vec{p}}{(2\pi)^2}\, \frac{e^{i\vec{p}\cdot\vec{r}}}{\xi_\vec{p}^2+\kappa^2\vec{p}^2 - E^2 } =-\frac{2\nu}{\Delta_t\sqrt{1-(E/\Delta_t)^2}}
\text{Im} \Big\{K_0[-i(1+i\tilde{\kappa})k_Fr] \Big\},\\
X_1(\vr{r}) &= - \intt \frac{d\vec{p}}{(2\pi)^2}\, \frac{\xi_p e^{i\vec{p}\cdot\vec{r}}}{\xi_\vec{p}^2+\kappa^2\vec{p}^2 - E^2}=-2\nu
\text{Im} \Big\{ \big(1+i\frac{\tilde{\kappa}}{\sqrt{1-(E/\Delta_t)^2}}\big)K_0[-i(1+i\tilde{\kappa})k_Fr] \Big\},\\
X_2^\pm(\vr{r}) &= \pm \intt \frac{d\vec{p}}{(2\pi)^2}\, \frac{i\kappa p_{\pm}e^{i\vec{p}\cdot\vec{r}}}{\xi_\vec{p}^2+\kappa^2\vec{p}^2 - E^2}=-\frac{2\nu}{\sqrt{1-(E/\Delta_t)^2}}\frac{x\pm iy}{r}\text{Re} \Big\{\big(1+i\tilde{\kappa}\sqrt{1-(E/\Delta_t)^2}\big)K_1[-i(1+i\tilde{\kappa})k_Fr] \Big\}.
\end{split}
\end{equation*}
In the above expressions the functions $K_\mu (x)$ stand for modified Bessel functions of the second kind. These expressions have been obtained by linearizing the bulk dispersion and are valid to the order $\mathcal{O} (\tilde{\kappa}^2)$. For large arguments the functions satisfy $X_{0/1}(\vr{r}), X_2^{\pm}(\vr{r})\sim \frac{e^{ik_Fr}}{\sqrt{k_Fr}}e^{-r/\xi_E}$ which means that hybrization of adjacent impurity states at midgap energies decay slowly for short distances and exponentially at distances longer than the coherence length $\xi_E^{-1}=\Delta_t\sqrt{1-(E/\Delta_t^2)^2}/v_F$. The spatial structure of the bound state wavefunctions imply that energy scale controlling the hybrization between two impurity states at distance $a$ apart is $\Delta_t/\sqrt{k_Fa}$.

For each impurity site $\vec{r}_i$, the  relation in Eq. (\ref{exact}) is satisfied. Therefore, for $N$ impurities, these equations form a closed set of equations for $2N$ subgap eigenvalues $E$ and eigenvectors $\Psi(\vec{r_i})=\begin{pmatrix} u(\vr{r}_i) \\ v(\vr{r}_i) \end{pmatrix}$ at the impurity sites. Instead of seeking an exact solution for the full range of parameters, we consider the eigenvalue equation for deep-lying eigenstates $|E|\ll \Delta_t$.  As discussed below, this approach will reproduce the exact topological phase diagram since the gap-closing transitions take place at $E=0$. In addition, we obtain the spectrum of the system in the deep-dilute  impurity regime $\beta\gg 1$, $\sqrt{k_Fa}\gg1$ where the spectrum is confined to the midgap region. Following Refs~[14, 16, 28], we can linearize the LHS of Eq. (\ref{exact}) with respect to $E$ and evaluate the coupling term on the RHS for $E=0$:
\begin{equation}\label{J}\tag{A.4}
\begin{gathered}
\Big[\mathbbm{1} + \beta\big(\frac{E}{\Delta_t}\tau_z - \gamma\mathbbm{1} \big)\Big] \Psi(\vr{r}_i)
= U\sum_{j\neq i} \lim_{\substack{E\to 0}} G_0(E,\vr{r}_i-\vr{r}_j)\,\tau_z\, \Psi(\vr{r}_j).
\end{gathered}
\end{equation}
Multiplying both sides by $\tau_z\beta^{-1}\Delta_t$ leads to 
\begin{equation*}
\begin{gathered}
\begin{pmatrix} \Delta_t(\beta^{-1}-\gamma)+E& 0 \\ 0 & -\Delta_t(\beta^{-1}-\gamma)+E\end{pmatrix} \Psi(\vr{r}_i)
=  \frac{\Delta_t}{\pi\nu}\sum_{j\neq i} \lim_{\substack{E\to 0}}  \begin{pmatrix} X_1(\vec{r}_{ij})& -iX_2^+(\vec{r}_{ij}) \\ iX_2^-(\vec{r}_{ij}) & -X_1(\vec{r}_{ij}) \end{pmatrix} \Psi(\vr{r}_j).
\end{gathered}
\end{equation*}
This equation can be written compactly as $\sum_jH_{ij}\Psi_j=E\Psi_i$, where $\Psi_i\equiv\Psi(\vec{r}_i)$ and $H_{mn} = \begin{pmatrix} h_{mn} & \Delta_{mn} \\ (\Delta_{mn})^\dagger & -h_{mn}^* \end{pmatrix}$,

\begin{equation}\label{final}\tag{A.5}
\begin{split}
h_{mn} &= \begin{dcases} \Delta_t(\gamma-\beta^{-1})& m=n \\
 \frac{\Delta_t}{\pi\nu}X_1(\vec{r}_{mn})_{|E=0} & m\neq n \end{dcases}
\\
\Delta_{mn} &= \begin{dcases} 0 & m=n \\
-i \frac{\Delta_t}{\pi\nu}X_2^+(\vec{r}_{mn})_{|E=0}& m\neq n. \end{dcases}
\end{split}
\end{equation}
This is equivalent to Eq.~(\ref{h2}) in the main text.

\subsection*{$H$ as a topological Hamiltonian}
The effective Hamiltonian (\ref{final}) is obtained through the steps outlined in Pientka \emph{et al.} in the pioneering work \cite{pientka2}. For deep impurities  $\beta\gg1$ that are weakly coupled $\sqrt{k_Fa}\gg1$ the bands are lying near the gap centre and the spectrum can be calculated by this approach. However, in the present case the effective Hamiltonian (\ref{final}) has more general utility beyond the low-energy theory. The expression (\ref{final}) can be regarded as a topological Hamitonian, providing access to the exact phase diagram of the full model (\ref{exact}) beyond the deep-dilute impurity limit. 

The role of the expression (\ref{final}) as topological Hamiltonian can be understood by the following arguments. The $E=0$ eigenstates of the exact problem (\ref{exact}) and effective problem $H\Psi=E\Psi$ coincide \emph{exactly}. This simply follows from the fact that the only approximation in the derivation of (\ref{final}) from  Eq.~(\ref{exact}) involves setting a number of energy arguments to zero,  a difference that does not  affect the $E=0$ solutions in any way. Since the topological phase transitions take place precisely when $E=0$ solutions exist, indicating the closing of the energy gap, the phase boundaries of the effective model and the full model necessarily coincide. In addition, in the vicinity of the phase boundaries the energy gap is small and the low-lying solutions of Eq.~(\ref{exact}) differ very little from the effective model and can be adiabatically deformed to each other effective. Therefore the Chern numbers calculated from the effective model (\ref{final}) match the Chern numbers of the full model, even in the regime where deep-dilute approximation is not applicable.  

Obtaining reliable energy eigenvalues is a different matter- those can be calculated from the model (\ref{final}) only for small energies $E/\Delta_t\ll1$ and with the accuracy that depends on the validity of the deep-dilute assumption. The corrections to the spectrum obtained from the effective Hamiltonian will be of the order of $\mathcal{O}(\frac{\beta^{-1}}{\sqrt{k_Fa}})$ which is second order in the small parameters in the deep-dilute regime. 

\subsection*{Including spin}
Here we discuss how our results of topological state engineering apply to the spinful chiral $p$-wave state with the $d$-vector of the triplet parametrization perpendicular to the plane. This has been the main candidate to describe superconductivity in strontium ruthenate. In addition to the Nambu matrices we introduce another set of Pauli matrices $\sigma_i$ and define $\sigma_0=\mathbb I_{2\times2}$.  In absence of potential impurities, the bulk is described by the $4\times4$ Hamiltonian
\begin{equation*}
\begin{gathered}
\mathcal{H}_\vec{p}^{(\rm bulk)} =  \xi_\vec{p}\tau_z\otimes\sigma_0 + \kappa\left( p_x\tau_x - p_y\tau_y \right)\otimes\sigma_z,
\end{gathered}
\end{equation*}
which is expressed in the basis $(\hat{\Psi}_{p\uparrow},\hat{\Psi}_{p\downarrow}, \hat{\Psi}^\dagger_{-p\downarrow}, -\hat{\Psi}^\dagger_{-p\uparrow})^T$. This state describes pairing of opposite spins in the orbital $L_z=1$ channel. The potential impurity term $\mathcal{H}^{(\rm imp)}(\vr{r}) =  U\tau_z\otimes\sigma_0\sum_n\,\delta(\vec{r}-\vec{r}_n)$ also has a diagonal spin structure. Thus by defining two spinors $\Psi_1=(\hat{\Psi}_{p\uparrow}, \hat{\Psi}^\dagger_{-p\downarrow})^T$, $\Psi_2=(\hat{\Psi}_{p\downarrow}, \hat{\Psi}^\dagger_{-p\uparrow})^T$, the full $4\times4$ BdG Hamiltonian can be transformed to two decoupled $2\times2$ blocks identical to the spinless model studied above. The impurity problem and topology can be studied separately for the blocks exactly as for spinless fermions. 

\subsection*{Effect of a non-localized potential}

Below we introduce a non-localized scalar potential to study the effect of other scattering channels than the s one. We make use 
of the formalism introduced in Ref. \cite{lutchyn2015}.

We consider a non-localized single potential impurity described by 
$
\mathcal{H}_{imp} = U(\mathbf{r}) \cdot \tau_z.
$
For that we decompose in momentum space $U(\mathbf{p}) = \sum\limits_{l}U_l(p) e^{il\theta}$. The unperturbed Green's function in momentum space reads:
$$
G(E,\mathbf{p}) = -\frac{1}{\xi_p^2+\varkappa^2 p^2 - E^2}
		\begin{pmatrix} 	E+\xi_p & i\varkappa p_+ \\ 
				-i\varkappa p_-  & E-\xi_p
		\end{pmatrix},
$$
Note that we omit the "0" index for the Green's function to stay consistent with Ref. \cite{lutchyn2015}. Rewriting the function above using the harmonic decomposition $G(E,\mathbf{p}) = \sum\limits_{n}G_n(E,p)e^{in\theta}$, where $p = |\mathbf{p}|$ we get:
\begin{align*}
G_0(E,p) &= -\frac{1}{\xi_p^2+\varkappa^2 p^2 - E^2}
		\begin{pmatrix} 	E+\xi_p & 0 \\ 
				0  & E-\xi_p
		\end{pmatrix}, \\
G_{-1}(E,p) &= -\frac{1}{\xi_p^2+\varkappa^2 p^2 - E^2}
		\begin{pmatrix} 	0 & 0 \\ 
				-i\varkappa p  & 0
		\end{pmatrix}, \\
G_{1}(E,p) &= -\frac{1}{\xi_p^2+\varkappa^2 p^2 - E^2}
		\begin{pmatrix} 	0 & i\varkappa p \\ 
				0  & 0
		\end{pmatrix}.
\end{align*}
All the higher harmonics corresponding to $|n|>1$ are absent in the bare Green's function. For further calculations we need to compute the averaged values of these functions over momenta, namely $\overline{G_n(E)} = \int\limits_0^{+\infty} \frac{pdp}{2\pi} G_n(E,p)$:
\begin{align*}
\overline{G_0(E)} = 
		\begin{pmatrix} 	
				E X_0 + X_1 & 0 \\ 
				0  & E X_0 - X_1
		\end{pmatrix},\quad
\overline{G_{-1}(E)} = 
		\begin{pmatrix} 	
				0 & 0 \\
				-i(\varkappa p_F X_0 + \tilde{\varkappa} X_1)  & 0
		\end{pmatrix}, \quad
\overline{G_{1}(E)} = 
		\begin{pmatrix} 	
				0 & i(\varkappa p_F X_0 + \tilde{\varkappa} X_1) \\ 
				0  & 0
		\end{pmatrix},
\end{align*}
with
$$
X_0 = -\frac{\pi \nu}{\sqrt{1+\tilde{\varkappa}^2}} \frac{1}{\sqrt{\Delta_t^2-E^2}}, \quad
X_1 = \frac{\pi \nu}{\sqrt{1+\tilde{\varkappa}^2}} \frac{\gamma\Delta_t}{\sqrt{\Delta_t^2-E^2}}, \quad 
\gamma = \frac{\tilde{\varkappa}}{\sqrt{1+\tilde{\varkappa}^2}}, \quad 
\Delta_t = \frac{\varkappa p_F}{\sqrt{1+\tilde{\varkappa}^2}},
$$
which have been calculated explicitly in Ref. \cite{kaladzhyan2016}.
For a bound state solution to exist, the following condition must be satisfied (Eq.~(S25) from Ref. \cite{lutchyn2015}):
\begin{equation*}
\det\left( \begin{array}{ccc}
\overline{G_0(E)} U_{-1} \tau_z-\tau_0 & \overline{G_{-1}(E)}U_{0} \tau_z & 0 \\
\overline{G_1(E)} U_{-1} \tau_z & \overline{G_{0}(E)} U_{0} \tau_z -\tau_0 & \overline{G_{-1}(E) }U_{1} \tau_z\\
0 & \overline{G_1(E)} U_{0} \tau_z & \overline{G_0(E)} U_{1} \tau_z-\tau_0
\end{array}\right)=0.
\end{equation*}
The equation above yields the energy levels. We next consider only the harmonics $l=\pm 1$ and $l=0$ for the scattering potential. This is enough for our purpose to demonstrate that they lead to Shiba states with energy near the gap edges. 
Note that the symmetry of the problem requires $U_{-1} = U_{1}$, and thus we get
$$
\frac{\beta_1 (\epsilon-\gamma)}{\sqrt{1-\epsilon^2}} = -1, \quad \frac{(\beta_0-\beta_1) \epsilon- \gamma(\beta_0+\beta_1)}{\sqrt{1-\epsilon^2}} = -(1+\beta_0\beta_1)
$$
where we denote 
$
\beta_{0,1} \equiv \frac{\pi \nu U_{0,1}}{\sqrt{1+\tilde{\varkappa}^2}}, \quad \epsilon \equiv \frac{E}{\Delta_t}.
$
Note that each of these equations is giving a positive-energy solution, which always has a particle-hole symmetric negative-energy partner as required by particle-hole symmetry. We solve these equations considering $\beta_0 > \beta_1$. This is a valid approximation since we expect the scatterings in the other channels than s to be weaker. Therefore, we obtain: 
$$
E^{\pm}_1 = \pm\frac{\gamma \beta_1^2 - \sqrt{1+\beta_1^2 (1-\gamma^2)}}{1+\beta_1^2} \Delta_t, 
\quad
E^{\pm}_2 = \pm\frac{\gamma (\beta_0^2-\beta_1^2) - (1+\beta_0\beta_1)\sqrt{1+\beta_0^2 + \beta_1^2 + \beta_0^2 \beta_1^2 - \gamma^2(\beta_0+\beta_1)^2}}{1+\beta_0^2 + \beta_1^2 + \beta_0^2 \beta_1^2} \Delta_t. 
$$
For $\beta_1 \to 0$ we expect these solutions to coincide with the ones we had before for a delta-like potential (in other words, when we take into account only the $s$-scattering channel). Indeed,
$$
\lim\limits_{\beta_1 \to 0} \left[ \mp \frac{\gamma \beta_1^2 - \sqrt{1+\beta_1^2 (1-\gamma^2)}}{1+\beta_1^2} \Delta_t \right] = \pm \Delta_t,
$$
and therefore two subgap states merge with the quasiparticle continuum. The other two
$$
\lim\limits_{\beta_1 \to 0} \left[ \mp \frac{\gamma (\beta_0^2-\beta_1^2) - (1+\beta_0\beta_1)\sqrt{1+\beta_0^2 + \beta_1^2 + \beta_0^2 \beta_1^2 - \gamma^2(\beta_0+\beta_1)^2}}{1+\beta_0^2 + \beta_1^2 + \beta_0^2 \beta_1^2} \Delta_t \right] = \mp \frac{\gamma \beta_0^2 - \sqrt{1+\beta_0^2 (1-\gamma^2)}}{1+\beta_0^2} \Delta_t 
$$
coincide with the ones we obtained previously for a fully localized impurity. 

The calculations above show that the states appearing due to other scattering channels are situated very close to the superconducting gap and, therefore, can be disregarded.
Therefore in the $s$-dominated scattering channel, we can prove that the low-energy Shiba states (which are the ones we keep as a low-energy basis) come from the delta-like potential approximation.

\subsection*{Spectrum of an infinite strip}

\begin{figure}
\includegraphics[width=0.8\columnwidth]{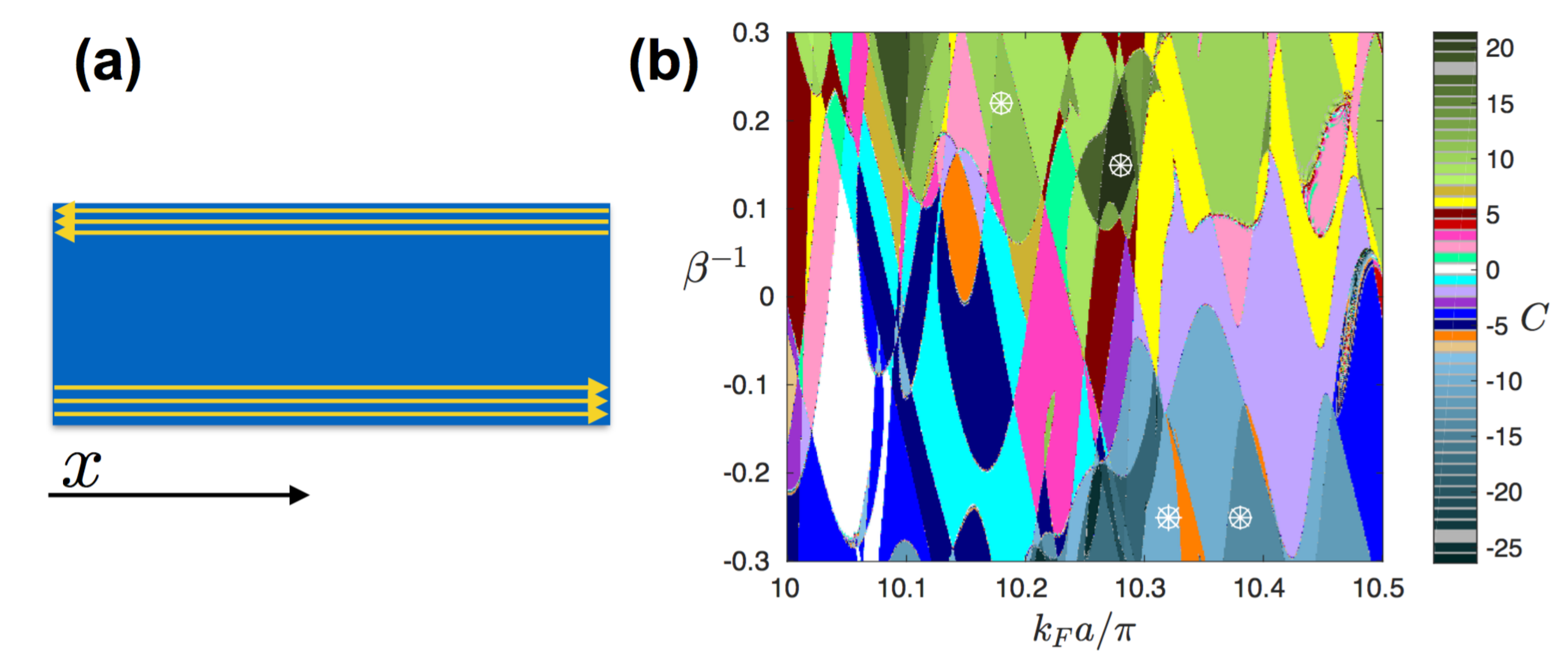} 
\caption{(a):In the strip geometry the boundaries support chiral edge modes that propagate in the opposite directions. (b): The phase diagram of the infinite square lattice with a lattice constant $a$ and coherence length $\xi/a=5$. The four points mark the parameter values employed in the Fig.~\ref{app2}.  }\label{app1}
\end{figure}
\begin{figure}
\includegraphics[width=0.99\columnwidth]{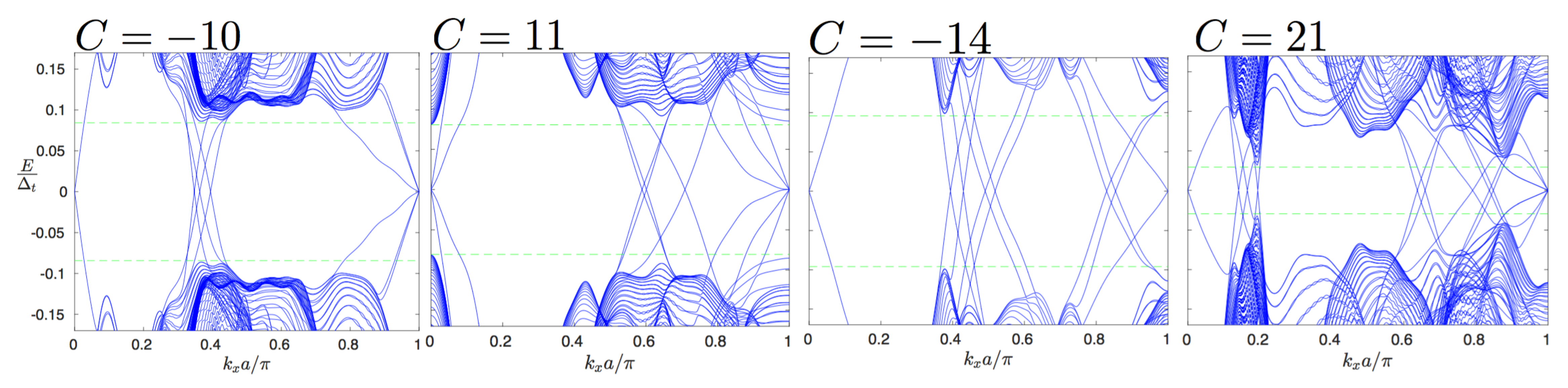} 
\caption{Spectrum of a strip as a function of momentum in the translation invariant direction in the positive half of the Brillouin zone. The spectra are reflection symmetric $k_x\to-k_x$ in the negative half. The different figures correspond to parameters indicated in the Fig.~\ref{app1}. The horizontal dashed line indicate the position of the bulk gap edge. }\label{app2}
\end{figure}

Here we illustrate topological properties of a system with finite width but infinite length. As depicted in Fig.~\ref{app1}, the chiral edge states are localized near the sample edge.  We assume a square lattice geometry with lattice constant $a$ and Fourier transform the Hamiltonian in $x$ direction. In Fig.~\ref{app2} we have plotted four sample spectra as a function of momentum $k_x$ corresponding to different Chern numbers. The edge state manifest as states traversing the bulk gap. Both edges support $|C|$ chiral edge states. 
Since the edge states plotted in Fig.~\ref{app2} have monotonic dispersions, each horizontal line in the bulk gap crosses $|C|$ states with positive slope and negative slope in the full Brillouin zone. These states are localized at the opposite edges. 

\end{document}